%% file: main.tex
\documentclass[%
 reprint,
superscriptaddress,
nobibnotes,
 amsmath,amssymb,
 aps,
 prl,
]{revtex4-2}

\usepackage{graphicx}
\usepackage{dcolumn}
\usepackage{bm}
\usepackage{hyperref}
\usepackage{siunitx}
\usepackage[capitalize]{cleveref}
\usepackage{physics}
\usepackage{amssymb}
\newcommand{\PT}{$\mathcal{PT}$}

\begin{document}

\title{Enhanced Sensitivity near a Quantum Exceptional Point \\in the Absence of Engineered Dissipation}

\def\RLEaffil{Research Laboratory of Electronics, Massachusetts Institute of Technology, Cambridge, MA 02139, USA}
\def\LLaffil{MIT Lincoln Laboratory, Lexington, MA 02421, USA}
\def\Physaffil{Department of Physics, Massachusetts Institute of Technology, Cambridge, MA 02139, USA}
\def\EECSaffil{Department of Electrical Engineering and Computer Science, Massachusetts Institute of Technology, Cambridge, MA 02139, USA}
\def\affileq{These authors contributed equally.}

\author{Réouven Assouly}
\altaffiliation{These authors contributed equally to this work.}
\affiliation{\RLEaffil} 

\author{Harry Hanlim Kang}
\altaffiliation{These authors contributed equally to this work.}
\affiliation{\RLEaffil}
\affiliation{\EECSaffil}

\author{Aziza Almanakly}
\affiliation{\RLEaffil}
\affiliation{\EECSaffil}

\author{Michael A. Gingras}
\author{Bethany M. Niedzielski}
\author{Hannah Stickler}
\author{Mollie E. Schwartz}
\affiliation{\LLaffil}

\author{Kyle Serniak}
\affiliation{\RLEaffil}
\affiliation{\LLaffil}

\author{Max Hays}
\affiliation{\RLEaffil}

\author{Jeffrey A. Grover}
\affiliation{\RLEaffil}
 
\author{William D. Oliver}
\email{william.oliver@mit.edu}
\affiliation{\RLEaffil}
\affiliation{\EECSaffil}
\affiliation{\Physaffil}

\date{June 12, 2026}

\begin{abstract}
Non-Hermitian systems exhibit phenomena absent from Hermitian systems, including exceptional points (EPs), at which two or more eigenvectors coalesce. Conventional implementations rely on gain and loss, which strongly limit quantum coherence. Here, following a proposal by Wang and Clerk~\cite{Wang19}, we realize a closed four-mode quantum system that emulates the dynamics of a \PT\ dimer---two coupled resonators with balanced gain and loss---without engineered dissipation. The four modes are implemented as harmonics of a superconducting coplanar-waveguide resonator, with parametric couplings engineered using a current-pumped SNAIL. We use this device as a sensor for small variations in the \PT \ dimer coupling strength. From signal-to-noise-ratio measurements, we observe enhanced sensitivity near the EP in a non-quantum-limited regime.
\end{abstract}

\maketitle

\paragraph{Introduction}
One of the foundational axioms of quantum mechanics is that observables are represented by Hermitian operators acting on a Hilbert space. In particular, the Hamiltonian—the operator associated with energy—must be Hermitian. Relaxing this requirement generally leads to complex energy eigenvalues, whose imaginary parts correspond to gain or loss. However, there exists a class of non-Hermitian Hamiltonians that can exhibit entirely real spectra, namely parity-time (\PT)-symmetric Hamiltonians~\cite{Bender1998, ElGanainy18, Miri19, Ashida20}.

\PT-symmetric systems are invariant under the combined action of the parity and the time-reversal operators. Consequently, their spectra are symmetric under complex conjugation: eigenvalues are either entirely real, corresponding to an unbroken \PT-symmetric phase, or appear as complex conjugate pairs in the spontaneously broken phase. The transition point between these two phases can be an exceptional point (EP), at which both eigenvalues and eigenvectors coalesce and the Hamiltonian becomes non-diagonalizable. This gives rise to a range of phenomena in the vicinity of an EP, including, for example, chiral mode switching~\cite{Doppler2016, Uzdin2011, Gilary2013, Milburn2015}.

One proposed application of non-Hermitian systems operating near an EP is sensing, as eigenvalues exhibit enhanced splitting in response to perturbations compared with conventional Hermitian systems~\cite{Wiersig14, Wiersig16, Liu16, Hodaei17, Chen17, Hajizadegan19, Farhat20, Rosa21, Sakhdari22, Wang22, Li23, Tang23}. This enhancement originates from the coalescence of eigenvectors at the EP. However, both theoretical studies~\cite{Lau18, Langbein18, Chen19, Duggan22, Ding23, Naikoo23, Loughlin24} and experimental demonstrations~\cite{Wang20, Almanakly25} have shown that enhanced eigenvalue sensitivity does not necessarily translate into an improved signal-to-noise ratio (SNR). In typical implementations, gain and loss processes amplify noise, thereby limiting achievable sensitivity.

In this work, we implement a system that circumvents this limitation by replacing incoherent gain and loss with coherent parametric drives. Specifically, we realize a \PT\ dimer~\cite{Guo09, Ruter10}, one of the simplest tunable \PT-symmetric systems, using a quantum analog simulator comprising four bosonic modes with engineered parametric interactions, as proposed in Ref.~\cite{Wang19}. We note that the single-mode version of the proposal has already been demonstrated with a Josephson parametric amplifier~\cite{Gaikwad23}. We demonstrate enhanced sensitivity to perturbations in the coupling strength near the EP, in a regime where quantum noise is not the dominant noise source.

\paragraph{Theory}
A \PT\ dimer consists of two resonant modes with amplitudes $\alpha_1(t)$ and $\alpha_2(t)$, subject to effective gain and loss rates $\gamma/2$ and $-\gamma/2$, respectively, and coupled with strength $g$. Its equations of motion can be written as
\begin{equation}
i\dv{t}\!
\begin{pmatrix}
\alpha_1 \\
\alpha_2
\end{pmatrix}
=\mathcal{H}_\mathrm{PT}
\begin{pmatrix}
\alpha_1 \\
\alpha_2
\end{pmatrix},
\end{equation}
where 

\begin{equation}
\mathcal{H}_\mathrm{PT} =
\begin{pmatrix}
 i\gamma/2 & g \\
g & -i\gamma/2
\end{pmatrix}
\end{equation}
is the \PT\ dimer Hamiltonian.
In the undercoupled (i.e. spontaneously broken) regime ($g<\gamma/2$) the eigenvalues are two complex-valued conjugate numbers, whereas in the overcoupled (i.e. \PT-symmetric) regime ($g>\gamma/2$) the eigenvalues are purely real. The transition occurs at $g=\gamma/2$, where the system reaches an EP.

The eigenvalues of the system are
\begin{equation}
    \omega_\pm = \pm \sqrt{g^2 - \left(\frac{\gamma}{2}\right)^2}.
\end{equation}
At the exceptional point, both the eigenvalues and the corresponding eigenvectors coalesce. A small perturbation of $\epsilon$ in $g$ then leads to an eigenvalue splitting
\begin{equation}
|\omega_+ - \omega_-| \sim 2\sqrt{\gamma |\epsilon|},
\end{equation}
which exhibits a square-root dependence on $\epsilon$, in contrast to the linear response characteristic of conventional Hermitian systems.

We emulate the non-Hermitian Hamiltonian $\mathcal{H}_\mathrm{PT}$ by mapping it onto a system of four bosonic modes, following Ref.~\cite{Wang19}. Specifically, we consider four modes labeled $\hat{a}_1$ through $\hat{a}_4$, as described in the following section. By applying four parametric drives, the Hamiltonian of the system takes the form
\begin{eqnarray}
\label{eq:full-ham}
\hat{H} &= &
g 
\left(
\hat{a}_1^\dagger \hat{a}_2
-
\hat{a}_3^\dagger \hat{a}_4
\right) \nonumber \\
~~~~~~ && +~
i \frac{\gamma}{2}
\left(
\hat{a}_1^\dagger \hat{a}_4^\dagger
-
\hat{a}_2^\dagger \hat{a}_3^\dagger
\right)
+ \mathrm{H.c.}
\end{eqnarray}

We then combine the bosonic mode amplitudes to reconstruct the effective \PT\ dimer modes $\alpha_1$ and $\alpha_2$, which we refer to as supermodes. In particular, using the quantum-mechanics free subsystem (QMFS) mapping~\cite{Tsang2012,Woolley2013,Didier2015,Moller2017,Khalili2018}
\begin{eqnarray}
\label{eq:mapping}
\alpha_1 &\leftrightarrow& \expval{\hat{a}_1 + \hat{a}^\dagger_4} \\
\alpha_2 &\leftrightarrow& \expval{\hat{a}_2 + \hat{a}^\dagger_3},
\end{eqnarray}
the resulting equations of motion for the supermodes reproduce those generated by $\mathcal{H}_\mathrm{PT}$.

\paragraph{Device and Calibration}
\begin{figure}[t!]
\includegraphics{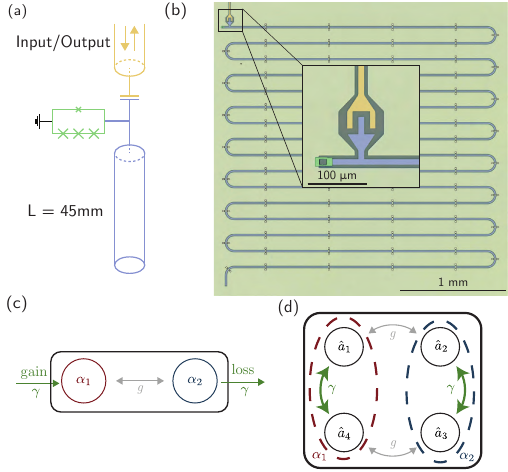}
\caption{\textbf{Device and experimental setup}. \textbf{(a)} Electrical schematic of the device: a \SI{45}{mm} transmission line (blue) shunted by a SNAIL (green) and capacitively coupled to a feedline (yellow). \textbf{(b)} False-color micrograph of the device, with a zoomed-in view of the capacitor and SNAIL in the inset. \textbf{(c)} Schematic of a \PT\ dimer. \textbf{(d)} Closed four-mode system onto which the \PT\ dimer is mapped. Gray arrows represent beam-splitting interactions; green arrows represent two-mode squeezing.}
\label{fig:schematic}
\end{figure}

Our system consists of a multimode superconducting resonator coupled to a SNAIL~\cite{Frattini17}. The latter consists of a small Josephson junction in parallel with three larger ones and provides the third order nonlinearity required for parametric mode swapping and two-mode squeezing~\cite{Bergeal10}.
The multimode resonator is a \SI{45}{mm}-long coplanar waveguide that is open at one end and shorted to ground through a SNAIL at the other. It supports many resonant modes with a fundamental $\lambda/4$ mode frequency of approximately \SI{580}{MHz}. For this experiment, we use four modes, $\hat{a}_1$, $\hat{a}_2$, $\hat{a}_3$, and $\hat{a}_4$, at frequencies near \SI{4.1}{GHz}, \SI{5.3}{GHz}, \SI{6.6}{GHz}, and \SI{7.8}{GHz}, respectively. The modes are unevenly spaced due to the finite inductance of the SNAIL.
As shown in \cref{fig:schematic}, a coupling port is placed near the shorted side of the resonator to enable pumping of the SNAIL and monitoring of the four modes. 

The SNAIL is flux-biased with an external coil such that the fourth-order nonlinearity in the system (Kerr) vanishes while three-wave mixing is retained~\cite{Sivak19}. This three-wave mixing enables parametric control of the mode-swapping and two-mode squeezing interaction strengths with microwave drives.

\begin{figure}[t!]
\includegraphics{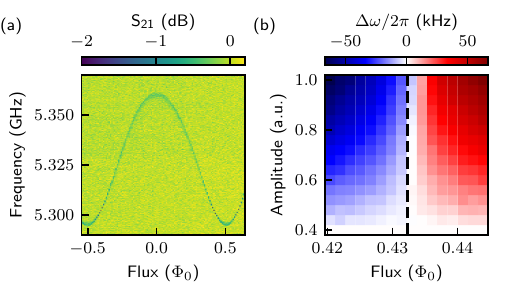}
\caption{\textbf{Basic characterization of mode $\hat{a}_2$}. \textbf{(a)} Mode frequency as a function of flux bias. \textbf{(b)} Frequency shift as a function of flux bias and excitation amplitude near the Kerr-free point. At zero flux, the resonant frequency shifts downward with increasing power, indicating a negative Kerr coefficient. At half flux, or $0.5\Phi_0$, the Kerr coefficient is positive; near a flux of $0.432\Phi_0$, the Kerr response cancels.
}
\label{fig:spectro}
\end{figure}

\cref{fig:spectro} shows the basic characterization of mode $\hat{a}_5$. We measure its frequency as a function of flux bias by fitting the spectroscopy data.
We then measure and fit its resonance frequency as a function of drive amplitude and external flux. The shift in mode frequency with drive amplitude changes sign near the Kerr-free flux-bias point. At the Kerr-free point, the mode frequencies are independent of drive power up to corrections from higher-order terms. 

\begin{figure}[t!]
\includegraphics{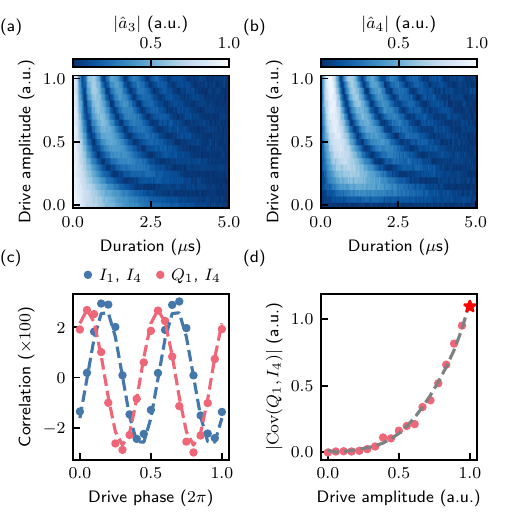}
\caption{\textbf{Calibration of parametric interactions}. \textbf{(a),(b)} Measured amplitudes of modes $\hat{a}_3$ and $\hat{a}_4$, respectively, as functions of the amplitude and duration of the mode-swapping pump $g_{34}$ when the system is initialized with a coherent state in mode $\hat{a}_3$. \textbf{(c)} Correlation between single-shot measurements of the quadratures of modes $\hat{a}_1$ and $\hat{a}_3$ as a function of the two-mode squeezing pump phase $\gamma_{14}$. The correlations are fitted with a cosine. \textbf{(d)} Amplitude of the covariance oscillations as a function of drive amplitude. The correlation is fitted with a full Gaussian model with two fit parameters, yielding a two-mode squeezing rate of \SI{118}{kHz/V} (See Supplementary Material). The red star corresponds to the data shown in (c).}
\label{fig:cal}
\end{figure}

Next, we calibrate the parametric drives. For mode swapping, we drive the system at the difference frequency of the two modes of interest (\cref{fig:cal}). We observe parametrically activated swapping, i.e., coherent population oscillations, between the two modes at a rate proportional to the pump amplitude. We target identical mode-swapping rates between modes 1 and 2 and between modes 3 and 4 by adjusting the relative pump amplitudes.

For two-mode squeezing, we drive at the sum frequency of the two modes. We confirm the presence of two-mode squeezing by measuring correlations between the two modes, as shown in \cref{fig:cal}(c). As expected, the phase of those correlations oscillates as a function of pump phase, and their magnitude increases with parametric-drive amplitude (\cref{fig:cal}(d)).
The relative phases of the parametric interactions are also calibrated according to \cref{eq:full-ham}.

\begin{figure}[t!]
\includegraphics[width=0.48\textwidth]{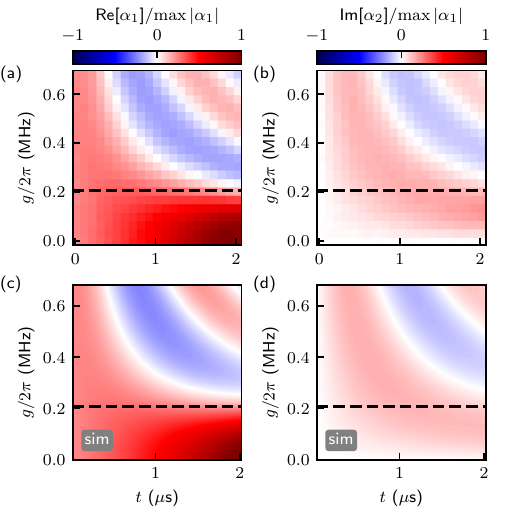}
\caption{
\label{fig:supermodes}
\textbf{Time-domain measurements of the normalized supermodes $\alpha_1$ and $\alpha_2$}. \textbf{(a)} The real part of the normalized supermode $\alpha_1$ as a function of the mode-swapping coupling rate $g$ and the pulse duration $t$. The supermodes are normalized such that the maximum of the absolute value of the amplitude is one.
The dashed line represents the location of the EP: $g=\gamma/2$.
\textbf{(b)} The imaginary part of the supermode $\alpha_2$. For better visual comparison, $\alpha_2$ is normalized by the same factor as $\alpha_1$.
\textbf{(c)} Fit to \textbf{(a)}. The simulation model incorporates mode loss rates $\Gamma_1$. The two-mode-squeezing rates, $\gamma_{14}$ and $\gamma_{23}$, along with the readout efficiencies of the two supermodes, are treated as fit parameters. The model agrees well with the experimental data.
\textbf{(d)} Fit to \textbf{(b)}, using the same model and fitting procedure.}
\end{figure}

\paragraph{Results}
Finally, we measure the time evolution of the emulated \PT\ dimer at the Kerr-free operating point under parametric drives, as shown in \cref{fig:supermodes}. In this experiment, the coupling strength $g$ and the parametric-drive duration $t$ are varied, while the gain/loss rate $\gamma$ is held fixed. The supermode amplitudes are reconstructed by combining the measured mode amplitudes $(\expval{\hat{a}_1}, \expval{\hat{a}_4})$ and $(\expval{\hat{a}_2}, \expval{\hat{a}_3})$ to calculate the supermodes $\alpha_1$ and $\alpha_2$, respectively, as defined in \cref{eq:mapping}. 
Although the platform enables simulation of a general detuned \PT\ dimer, we focus here on the zero-detuning case relevant for sensing applications. The system is initialized in the effective-gain supermode $\alpha_1$ by equally populating modes 4 and 7. Subsequent dynamics reveal population transfer to the effective-loss supermode $\alpha_2$ mediated by mode-swapping interactions.

Two distinct dynamical regimes emerge depending on the relative magnitudes of $g$ and $\gamma$. For $g < \gamma/2$, both $\alpha_1$ and $\alpha_2$ exhibit exponential growth, consistent with imaginary eigenvalues. In this regime, even the nominally lossy mode is amplified by continuous population transfer from the effective-gain mode, although its amplitude remains smaller. In contrast, for $g > \gamma/2$, the system exhibits coherent oscillations between $\alpha_1$ and $\alpha_2$, forming a characteristic Rabi-chevron pattern associated with real eigenvalues. The transition between these regimes occurs at the exceptional point, indicated by the dashed line.

The measured dynamics are in qualitative agreement with numerical simulations that incorporate experimentally observed nonidealities. We identify two main nonidealities: a finite mode lifetime and an asymmetry in the two-mode-squeezing rates. The first one is expected from the design since we engineered a small coupling to the drive line in order to be able to measure the modes. We attribute the second effect to imperfect initialization of the four modes. In practice, driving one mode leads to residual off-resonant excitation of the others, resulting in imperfect control of the initial populations of each resonator mode. Since two-mode squeezing amplifies each mode in proportion to the population of its partner, such imbalances lead to systematic errors in calibrating the two-mode-squeezing rates. This effect is included in the fitted model, yielding a ratio $\gamma_{14}/\gamma_{23} \approx 0.67$.

Nevertheless, we observe enhanced sensitivity to perturbations in the coupling strength in the vicinity of the exceptional point (EP), $g=\gamma/2$, where we take the average two-mode squeezing rate $\gamma=(\gamma_{14}+\gamma_{23})/2$ from the fit. Following Ref.~\cite{Almanakly25}, we define the normalized coupling strength $\tilde{g}\equiv 2g/\gamma$ such that at the EP, $\tilde{g}=1$. We define the sensitivity as

\begin{equation}
\eta_i(\tilde{g},t) \equiv \abs{\pdv{\alpha_i(\tilde{g},t)}{\tilde{g}}} \frac{1}{\sigma_i(\tilde{g},t)},
\end{equation}
where $i=1,2$ labels the supermodes and $\sigma_i(\tilde{g},t)$ denotes the standard deviation of the averaged measured signal. For each value of $\tilde{g}$, we choose the pulse duration $t$ that gives the best SNR, such that

\begin{eqnarray}
\eta_i (\tilde{g})&\equiv&\max_t\eta_i(\tilde{g},t)\nonumber \\ &=&\max_t \abs{\pdv{\alpha_i(\tilde{g},t)}{\tilde{g}}} \frac{1}{\sigma_i(\tilde{g},t)}.
\end{eqnarray}

\begin{figure}
\includegraphics[width=0.48\textwidth]{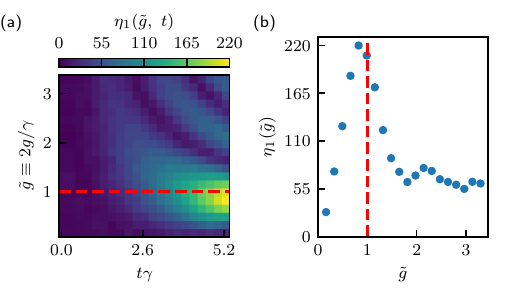}
\caption{\textbf{Sensitivity near the quantum exceptional point}. \textbf{(a)} Sensitivity to perturbations in the coupling $\tilde g$, $\eta_1$, extracted from the measurement $\alpha_1$ as a function of the beam-splitter coupling rate $g$ and pulse duration $t$. \textbf{(b)} Maximum sensitivity $\eta_1$ obtained from $\alpha_1$ as a function of $g$. The dashed line indicates the location of the exceptional point. Enhanced sensitivity is observed in its vicinity.}
\label{fig:sensitivity}
\end{figure}

The results are shown in \cref{fig:sensitivity}. We observe enhanced sensitivity near the exceptional point. In our system, this enhancement primarily reflects the increased response of the supermode amplitude to variations in $g$, rather than a reduction in noise. The measurements are dominated by classical noise contributions that exceed quantum noise sources, such as resonator shot noise and vacuum fluctuations introduced by the parametric drives. This is consistent with our operating regime of low photon number and modest drive amplitudes, as well as the limited gain of the TWPA, which was configured to provide broadband amplification over the 4--7~GHz range required for simultaneous readout of all four modes. As a result, the sensitivity is effectively proportional to the signal response, $\abs{\pdv*{\alpha_i}{\tilde g}}$, which is maximized near the exceptional point where the eigenvalues exhibit the strongest dependence on $\tilde g$.

\paragraph{Conclusion}
In summary, we realize a \PT\ dimer using a parametrically driven superconducting circuit comprising four bosonic modes. The implementation does not rely on incoherent gain or loss; such processes enter only as parasitic effects rather than defining elements of the system Hamiltonian. We demonstrate that multiple beam-splitting and two-mode-squeezing drives can be operated simultaneously and coherently. Within this platform, we observe the exceptional point and the corresponding enhancement in sensitivity to perturbations of system parameters. The observed behavior is consistent with a classical-noise-limited regime and with prior analyses of quantum-noise-limited settings.

Beyond sensing applications, the analog simulator developed here provides a versatile platform for exploring non-Hermitian physics. In particular, detuned \PT\ dimer dynamics can be accessed by introducing frequency detuning in the beam-splitting drives. Building on the present demonstration, future directions include exploring phenomena such as chiral mode switching.
\\
\paragraph{Acknowledgments}
The authors thank Shantanu Jha and Beatriz Yankelevich for fruitful discussions.
This material is based upon work supported in part by the Air Force Office of Scientific Research Multidisciplinary University Research Initiative (MURI) under award number FA9550-22-1-0166; in part by U.S. Army Research Office Grant No. W911NF-23-1-0045; and in part under Air Force Contract No. FA8702-15-D-0001.
HHK is supported by the Korea Foundation for Advanced Studies.
Any opinions, findings, conclusions, or recommendations expressed in this material are those of the author(s) and should not be interpreted as necessarily representing the official policies or endorsements of the U.S. Government.
\\
\paragraph{Author Contributions}
R.A. and H.H.K. conducted the measurements, performed theoretical calculations and simulations, analyzed data, and wrote the manuscript.
R.A. designed the experimental procedure and the device.
A.A. supported the data analysis.
M.G., B.M.N., and H.S. fabricated the devices with coordination from K.S. and M.E.S.
K.S., M.H., J.A.G., and W.D.O. supervised the project.
All authors discussed the results and commented on the manuscript.
\\
\paragraph{Code and Data Availability}
The code and data that support the findings of this study are available from the corresponding author upon reasonable request.

\bibliography{biblio}

\include{supp.tex}

\end{document}

%% file: supp.tex
\onecolumngrid
\newpage
\begin{center}
    \textbf{SUPPLEMENTARY INFORMATION}
\end{center}

\setcounter{figure}{0}
\setcounter{equation}{0}

\subsection{Device and Experimental Setup}
\begin{figure}[b]
    \centering
    \includegraphics[width=\linewidth]{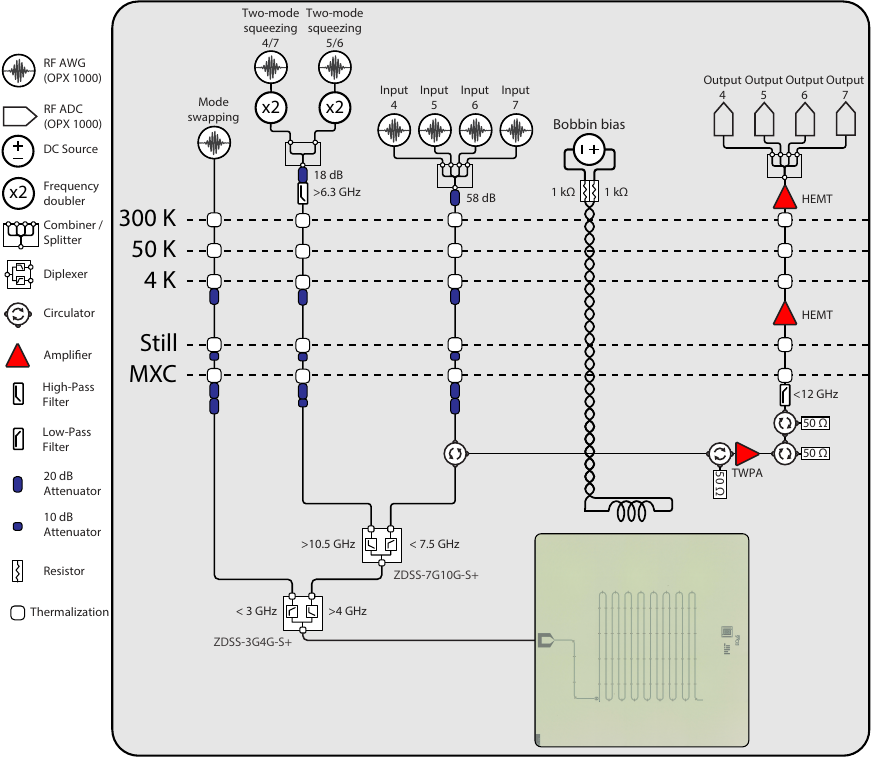}
    \caption{\textbf{Experimental setup.} Wiring schematic of the cryogenic and room-temperature setup used for the measurements.}
    \label{fig:suppl_exp_setup}
\end{figure}
This experiment was performed in a Leiden Cryogenics CF-CS81 dilution refrigerator, which operates at a base temperature of approximately \SI{8}{mK}. The wiring schematic is shown in \cref{fig:suppl_exp_setup}. The device was fabricated and packaged at MIT Lincoln Laboratory in a copper- and gold-plated sample holder, with an external superconducting coil used to apply a magnetic field. The coil leads are filtered (not shown) at the \SI{4}{K} stage using an RC filter with a cutoff frequency of approximately \SI{30}{kHz}, and are connected to a Yokogawa 7651, operated in constant-voltage mode, through a pair of \SI{1}{k\ohm} resistors. The device is protected by three layers of shielding: an aluminum shield, a tin-plated copper shield, and a Cryoperm magnetic shield.

All signals are generated and measured using two MW-FEM modules of an OPX-1000 from Quantum Machines. Because of the high frequency of the two-mode-squeezing pump, a frequency doubler was required for the two-mode-squeezing lines. The lines were split into two paths to avoid intermodulation in the nonlinear doubler. The power response of the doubler was subsequently calibrated such that the x axis of panel (b) in \cref{fig:cal} shows the amplitude at the input of the device.

At the mixing chamber, two diplexers from Mini-Circuits are used to combine and separate the pumps and signals. The signals are then amplified by a traveling-wave parametric amplifier (TWPA) provided by Lincoln Laboratory, followed by further amplification using an LNF-LNC0.3\_14 HEMT at \SI{4}{K} and a MITEQ amplifier at room temperature. Because of the large frequency separation, the signal is then split into four paths and demodulated by four separate channels of the OPX.
The specific control and measurement equipment used throughout the experiment is summarized in Table~\ref{tab:equipment}.
The relevant parameters of the device used in the experiment are summarized in Table~\ref{tab:params}.

\begin{table}[h!]
\centering
\begin{tabular}{lll}
\hline
\hline
Component & Manufacturer & Model \\
\hline
Dilution Refrigerator & Leiden Cryogenics & CF-CS81 \\
DC Source & Yokogawa & 7651 \\
AWG \& ADC & Quantum Machines & OPX-1000 \\
\hline
\hline
\end{tabular}
\caption{\textbf{Summary of control equipment.} Manufacturers and model numbers of the experimental control equipment.}
\label{tab:equipment}
\end{table}

\begin{table}[h!]
\centering
\begin{tabular}{lllll} 
\hline
\hline
Parameter & $\hat{a}_1$ & $\hat{a}_2$ & $\hat{a}_3$ & $\hat{a}_4$ \\
\hline
Frequency (\si{\GHz}) & 4.07 & 5.30 & 6.55 & 7.80\\
$\Gamma_{1,k}/2\pi$ (\si{\kHz}) & 47 & 50 & 190 & 202\\
$\Gamma^L_{2,k}/2\pi$ (\si{\kHz}) & 43 & 55 & 72 & 105\\
$\Gamma^G_{2,k}/2\pi$ (\si{\kHz}) & 11 & 12 & 9 & 18\\
\hline
\hline
\end{tabular}
\caption{\textbf{Summary of device parameters.} Operational mode frequencies, energy loss rate $\Gamma_{1,k}$, and longitudinal and Gaussian dephasing rates $\Gamma^L_{2,k}$ and $\Gamma^G_{2,k}$.}
\label{tab:params}
\end{table}

\subsection{Individual Mode Calibration}
In this section, we discuss the calibration of the individual bosonic modes $\hat{a}_1$ through $\hat{a}_4$. This calibration depends strongly on several factors, including the TWPA pump frequencies and power, as well as filtering in the drive and readout lines.

\begin{figure*}[h]
    \centering
    \includegraphics[width=0.8\textwidth]{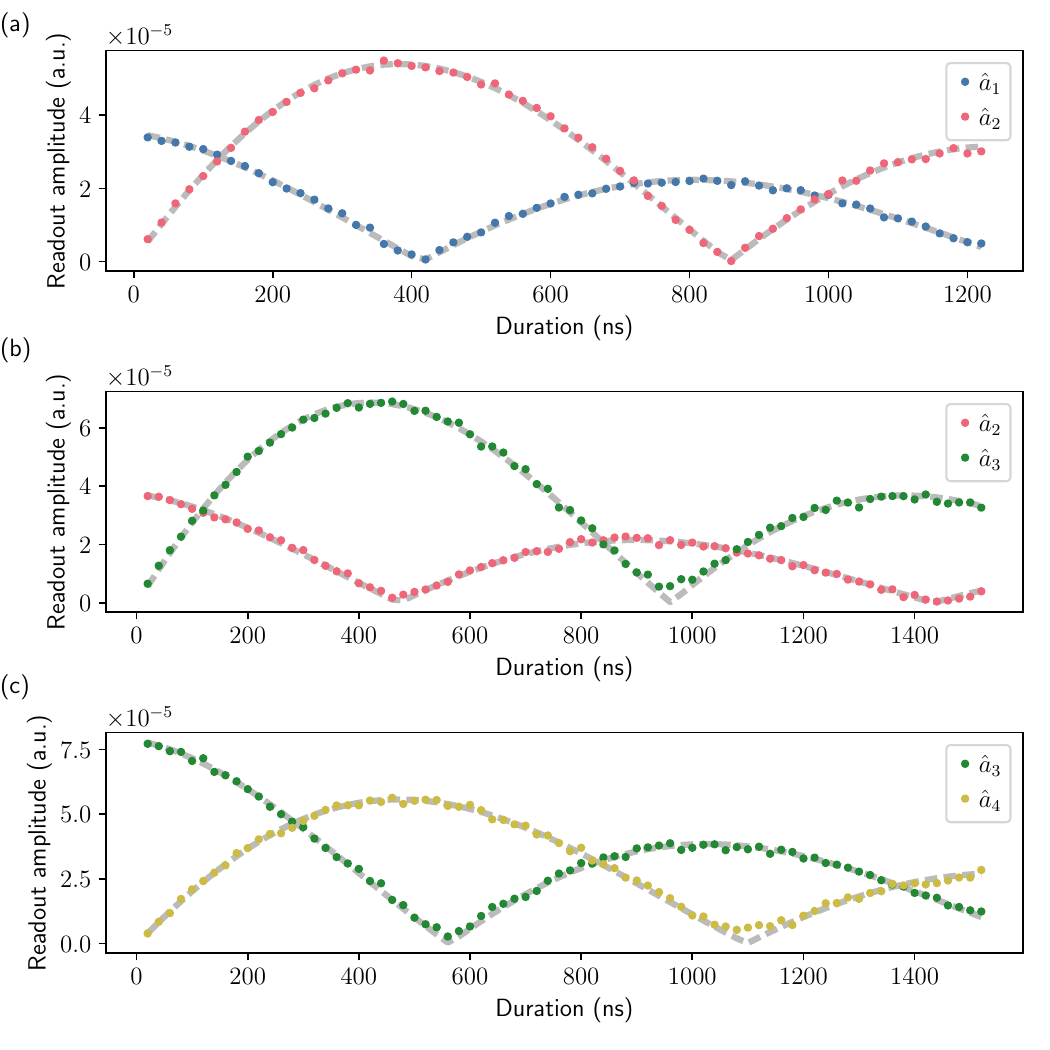}
    
    \caption{\textbf{Readout gain calibration.} Mode-swapping measurements for \textbf{(a)} modes $\hat{a}_1$ and $\hat{a}_2$, \textbf{(b)} modes $\hat{a}_2$ and $\hat{a}_3$, and \textbf{(c)} modes $\hat{a}_3$ and $\hat{a}_4$. Grey dotted lines denote exponentially decaying sinusoidal fits. The ratios of the fitted amplitudes are used to extract the relative readout efficiencies between the two modes.}
    \label{fig:suppl_readout_efficiency}
\end{figure*}

The first step is to extract the relative readout efficiencies of the four modes, $\tilde{\eta}_k$. We perform mode-swapping pulses between two modes of interest, measure the readout voltages as a function of pulse duration, and fit the data, as shown in Fig.~\ref{fig:suppl_readout_efficiency}. The fitting model is an exponentially decaying sinusoid. The ratio of the fitted amplitudes gives the relative readout gains.

\begin{figure*}[h]
    \centering
    \includegraphics[width=0.8\textwidth]{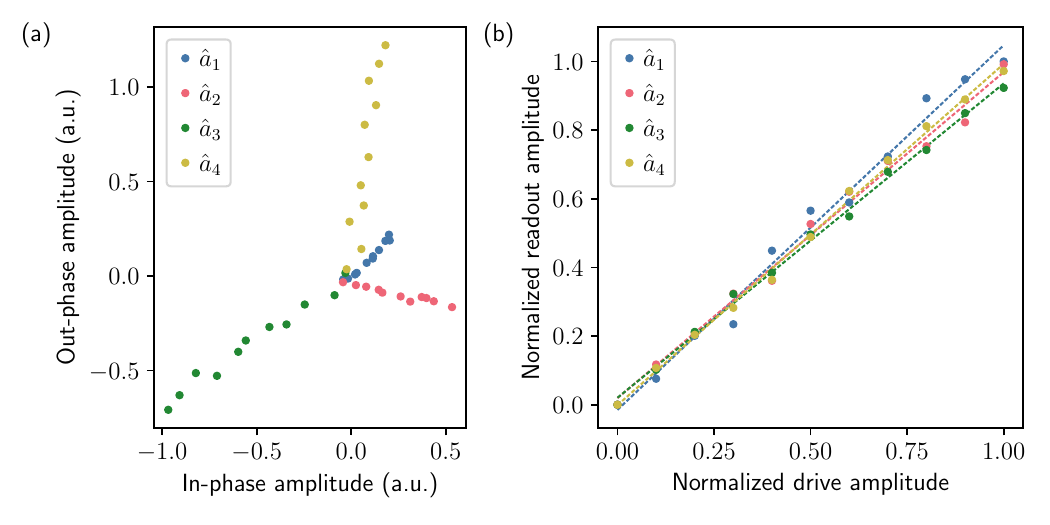}
    
    \caption{\textbf{Mode drive calibration.} \textbf{(a)} In-phase and quadrature readout amplitudes of all four modes, measured with varying drive amplitude and plotted in the IQ plane. \textbf{(b)} Normalized readout amplitude of all four modes as a function of normalized drive amplitude. The readout amplitudes are normalized by subtracting the readout offset of each mode and dividing by the relative readout efficiencies $\tilde{\eta}_k$. After calibrating $A_k$, applying the same normalized drive amplitude to all four modes leads to the same normalized readout amplitude.}
    \label{fig:suppl_mode_drive}
\end{figure*}

Next, we calibrate the individual mode drives. We apply mode drives for a fixed duration, vary the drive amplitude, and measure the in-phase and quadrature amplitudes $V_{I,k}$ and $V_{Q,k}$ at the end of the drive, simultaneously for four modes. As shown in Fig.~\ref{fig:suppl_mode_drive}, the heterodyne readout traces out a linear trajectory in the IQ plane. The angle associated with this linear trace defines the mode phase, $\theta_k$, which arises from several factors, including phase delays throughout the lines.

We also measure the readout offset $V_{0,k}$, defined as the readout voltage when no drive is applied. This offset arises from parasitic leakage of the local-oscillator tone into the readout port. Following this calibration, all other measurement results are post-processed as
\begin{equation}
\expval{\hat{a}_k} \equiv \exp(-i\theta_k)\left(\expval{V_{I,k} + iV_{Q,k}} - V_{0,k}\right)/\tilde{\eta}_k, 
\label{eq:a_k}
\end{equation}
such that the initial distribution of each mode populated by the mode drive is aligned along the in-phase axis. The corresponding value of $\expval{\hat{a}^\dagger_k}$ is then obtained by complex conjugation.

We then extract the drive ratio for all four modes, $A_k$, which rescales the mode-drive amplitude to achieve the same readout amplitude $\expval{\hat{a}_k}$. In general, this ratio is not unity because of the frequency dependence of the drive-line attenuation and the different coupling strengths to the four modes.

This ratio is used to achieve equal initial populations in the two bosonic modes that comprise a pseudomode.

\begin{figure*}[h]
    \centering
    \includegraphics[width=0.8\textwidth]{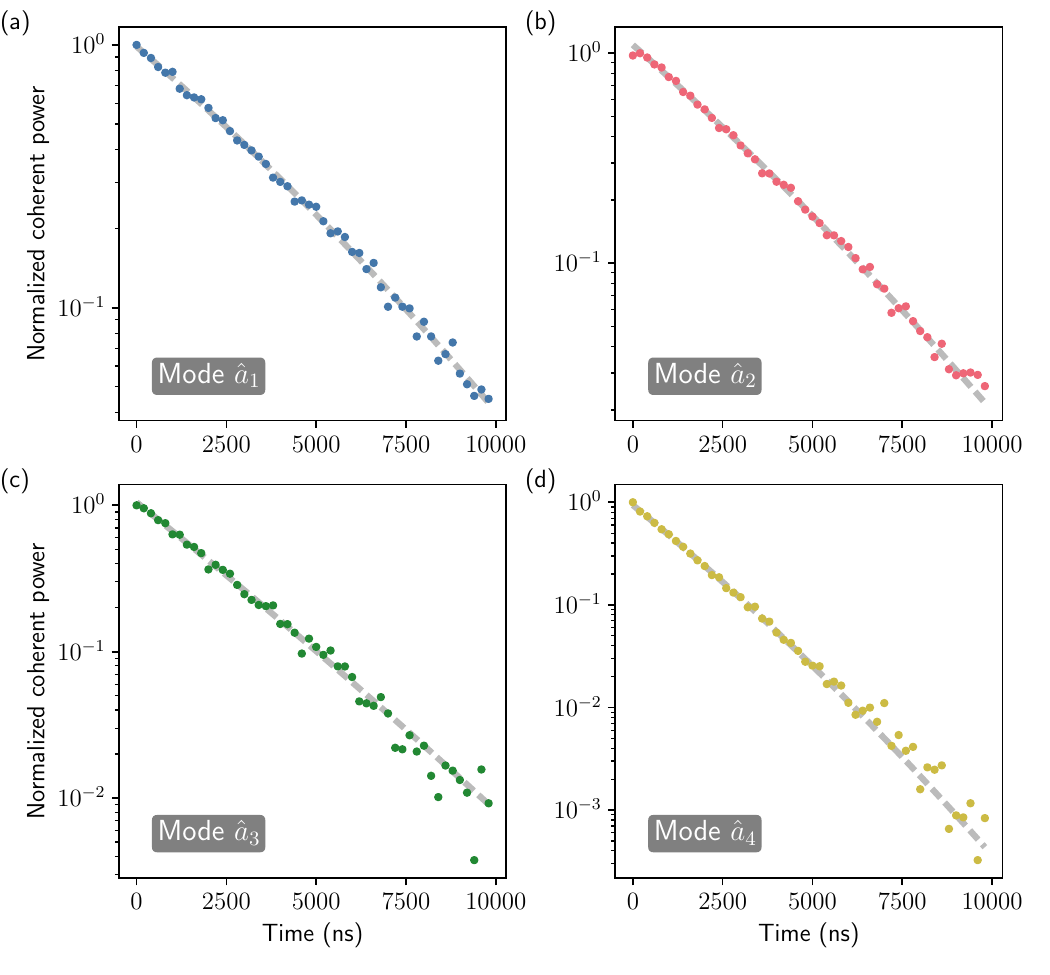}
    \caption{\textbf{Mode $\Gamma_2$ calibration.} Coherent power component of the ringdown readout signal for \textbf{(a)} mode $\hat{a}_1$, \textbf{(b)} mode $\hat{a}_2$, \textbf{(c)} mode $\hat{a}_3$, and \textbf{(d)} mode $\hat{a}_4$, plotted as a function of time. Grey dotted lines denote gaussian-plus-exponential fits using Eq.~\ref{eq:coh_ringdown}, from which the mode $\Gamma_2$ values are extracted.}
    \label{fig:suppl_coherent_ringdown}
\end{figure*}

\begin{figure*}[h]
    \centering
    \includegraphics[width=0.8\textwidth]{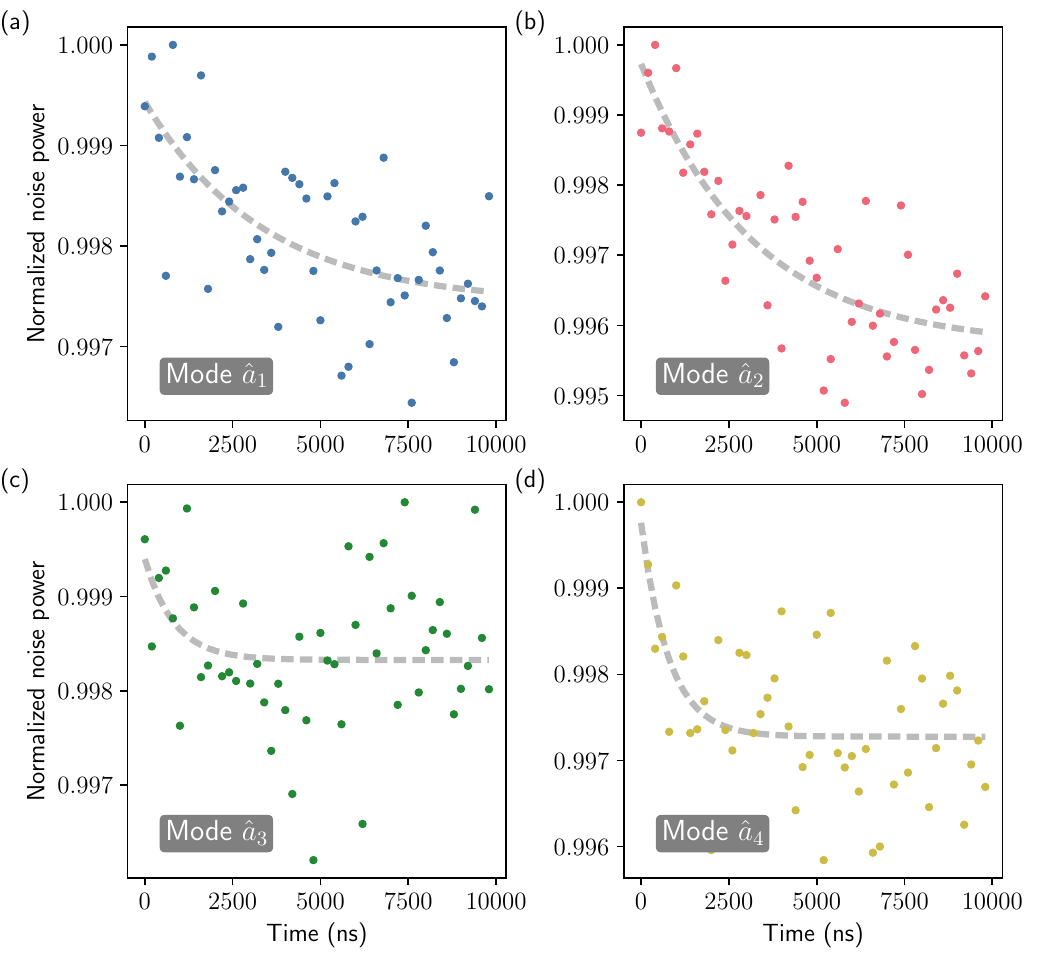}
    
    \caption{\textbf{Mode $\Gamma_1$ calibration.} Noise power component of the ringdown readout signal for \textbf{(a)} mode $\hat{a}_1$, \textbf{(b)} mode $\hat{a}_2$, \textbf{(c)} mode $\hat{a}_3$, and \textbf{(d)} mode $\hat{a}_4$, plotted as a function of time. Grey dotted lines denote exponential fits using Eq.~\ref{eq:coh_ringdown}, from which the mode $\Gamma_1$ values are extracted.}
    \label{fig:suppl_noise_ringdown}
\end{figure*}

Finally, we perform ringdown measurements to calibrate the mode lifetimes (Figs.~\ref{fig:suppl_coherent_ringdown},~\ref{fig:suppl_noise_ringdown}). After initially populating the four modes, we calculate the coherent and noise (incoherent) power components of the readout signal. The former is obtained by taking the squared magnitude of the mean signal amplitude, i.e., the power of the mean signal, while the latter is obtained by averaging the squared amplitudes of the individual single shots. We then perform Gaussian and exponential fits to extract $\Gamma_{2,k}=1/T_{2,k}$ and $\Gamma_{1,k}=1/T_{1,k}$ for the modes. Specifically, we fit to the model
\begin{eqnarray}
P_{\mathrm{coherent},k} &\propto& \exp \left(-\Gamma_{2,k}^L t - \left(\Gamma^G_{2,k} t\right)^2\right), \label{eq:coh_ringdown} \\
P_{\mathrm{noise},k} &\propto& \exp \left(-\Gamma_{1,k} t\right) \label{eq:noise_ringdown},
\end{eqnarray}
where $P_{\mathrm{coherent},k}$ and $P_{\mathrm{noise},k}$ are the coherent and noise powers of mode $\hat{a}_k$, respectively. The exponential component is dominated by white-noise-induced dephasing, while the Gaussian component is dominated by $1/f$ noise contributions to the dephasing rate $\Gamma_{2,k}$. The extracted rates are shown in Table~\ref{tab:params}.

\subsection{Two-mode squeezing modeling}
To model the evolution of the a pair of bosonic modes $(\hat a_1, \hat a_2)$ under a two-mode squeezing drive, we consider the system as a 2 mode gaussian state. This approximation is valid for our system since we only use linear displacements and quadratic operations such as mode swapping or two-mode squeezing. The main non-ideality, losses, also conserve the gaussian nature of the quantum state.

Let's define the quadratures of the two modes as 
\begin{equation}
    \hat X_i = \frac{\hat a_i + \hat a_i^\dagger}{2}, \hat P_i = \frac{\hat a_i - \hat a_i^\dagger}{2i}, 
\end{equation}
A two-mode gaussian state is entirely characterized by a 4-vector displacement $R = \left(X_1, P_1, X_2, P_2\right)^T$ and a 4x4 covariance matrix defined as
\begin{equation}
    V_{ij} = \frac{1}{2}\expval{R_i R_j + R_j R_i} - \expval{R_i}\expval{R_j}. 
\end{equation}

The two evolve independently from each other~\cite{Weedbrook2012}. In particular, the covariance follows the following equation
\begin{equation}
    \dot V = AV + VA^\dagger + D
    \label{eq:supp_gaussian_evol}
\end{equation}
with $A$ and $D$ the drift and diffusion matrices. Here, in the case of two  bosonic modes with loss rates $\Gamma_{1,1}, \Gamma_{1,2}$ coupled with a two-mode squeezing coupling $\gamma$, we have~\cite{Weedbrook2012}
\begin{equation}
    A = \begin{pmatrix}
        -\Gamma_{1,1}/2 & 0 & \gamma & 0\\
        0& -\Gamma_{1,1}/2 & 0 & -\gamma\\
        \gamma& 0 & -\Gamma_{1,2}/2 & 0\\
        0 & -\gamma & 0 & -\Gamma_{1,2}/2
    \end{pmatrix}
\end{equation}
and 
\begin{equation}
    D = \frac{1}{4}\begin{pmatrix}
        \Gamma_{1,1} & 0 & 0 & 0\\
        0 & \Gamma_{1,1} & 0 & 0\\
        0 & 0 & \Gamma_{1,2} & 0\\
        0 & 0 & 0 & \Gamma_{1,2}
    \end{pmatrix}.
\end{equation}

To fit \cref{fig:cal} (d), we numerically solve \cref{eq:supp_gaussian_evol} for $V_{1,3}$ at $t = \SI{2}{\micro s}$ with the assumption that the two-mode squeezing is proportional to the drive amplitude. The two free parameters are the readout gain (a scaling factor between the calculated dimensionless covariance and the measured covariance in \si{V^2}) as well as a scaling factor between the drive amplitude in \si{V} and the two-mode squeezing strength in \si{rad/s}. Note that due to the nonlinear response of the frequency doubler, before fitting the data, we map the x-axis that represents the drive amplitude before the doubler to the amplitude at the output of the doubler using the results from an independent calibration.

\subsection{Supermodes evolution modeling}
To fit the normalized supermode data in \cref{fig:supermodes}(a) and (b), we model the dynamics using mean-field equations of motion for four bosonic modes:
\begin{eqnarray}
\frac{d\expval{\hat{a}_1}}{dt}
&=& -\frac{\Gamma_{1,1}}{2}\expval{\hat{a}_1}
- i g \expval{\hat{a}_2}
+ \frac{\gamma_{14}}{2}\expval{\hat{a}^\dagger_4}, \\
\frac{d\expval{\hat{a}_2}}{dt}
&=& -\frac{\Gamma_{1,2}}{2}\expval{\hat{a}_2}
- i g \expval{\hat{a}_1}
- \frac{\gamma_{23}}{2}\expval{\hat{a}^\dagger_3}, \\
\frac{d\expval{\hat{a}_3}}{dt}
&=& -\frac{\Gamma_{1,3}}{2}\expval{\hat{a}_3}
+ i g \expval{\hat{a}_4}
+ \frac{\gamma_{23}}{2}\expval{\hat{a}^\dagger_2}, \\
\frac{d\expval{\hat{a}_4}}{dt}
&=& -\frac{\Gamma_{1,4}}{2}\expval{\hat{a}_4}
+ i g \expval{\hat{a}_3}
+ \frac{\gamma_{14}}{2}\expval{\hat{a}^\dagger_1},
\end{eqnarray}
and
\begin{eqnarray}
\alpha_1 &=& \tilde{\eta}_{\alpha,1} \expval{\hat{a}_1 + \hat{a}^\dagger_4}, \\
\alpha_2 &=& \tilde{\eta}_{\alpha,2} \expval{\hat{a}_2 + \hat{a}^\dagger_3}.
\end{eqnarray}
Here, $\expval{\hat{a}_k}$ denotes the post-processed mode amplitude obtained from the raw readout voltages using the mode phases $\theta_k$ and relative readout efficiencies $\tilde{\eta}_k$ according to Eq.~\ref{eq:a_k}.

The equations include the coherent dynamics described by the Hamiltonian in Eq.~\ref{eq:full-ham}, while also incorporating several experimentally relevant nonidealities.
First, each mode experiences energy decay at a rate $\Gamma_{1,k}$ due to coupling to the drive and readout feedline; these values are independently determined from ringdown measurements and listed in Table~\ref{tab:params}. Second, the two-mode-squeezing rates $\gamma_{14}$ and $\gamma_{23}$ are treated as independent fit parameters to account for imperfect initialization and calibration of the squeezing interactions. Third, the scale factors $\tilde{\eta}_{\alpha,1}$ and $\tilde{\eta}_{\alpha,2}$ allow for residual calibration errors in the relative amplitudes of the two measured supermodes.
Note that we also observe weak photon-number-dependent frequency shifts arising from residual higher-order Kerr nonlinearities. However, these shifts cancel to first order in the supermode observables and are therefore omitted from the model.

The fitted parameters are summarized in Table~\ref{tab:supermode_fit}.
\begin{table}[h!]
\centering
\begin{tabular}{ll}
\hline
\hline
Parameter & Value \\
\hline
$\gamma_{14}/2\pi$ (\si{\MHz}) & 0.33\\
$\gamma_{23}/2\pi$ (\si{\MHz}) & 0.50 \\
$|\tilde{\eta}_{\alpha,1}|$ & 0.23 \\
$|\tilde{\eta}_{\alpha,2}|$ & 0.14 \\
\hline
\hline
\end{tabular}
\caption{\textbf{Summary of fitting parameters for \cref{fig:supermodes}}. Two-mode-squeezing rates $\gamma_{14}$ and $\gamma_{23}$, and relative supermode amplitudes $\tilde{\eta}_{\alpha1}$ and $\tilde{\eta}_{\alpha2}$.}
\label{tab:supermode_fit}
\end{table}